\documentclass[floatfix,aps,a4paper,twocolumn, pra,showpacs,10pt]{revtex4-1}

\usepackage{graphicx}
\usepackage{amssymb}
\usepackage{amsmath}
\usepackage{color}
\usepackage{psfrag}
\usepackage{epsfig}
\usepackage{bbm}
\usepackage{bm}
\usepackage{dsfont}
\usepackage{hyperref}
\hypersetup{breaklinks}

\usepackage{xspace}

\newcommand{\ket}[1]{| #1 \rangle}
\newcommand{\bra}[1]{\langle #1 |}
\newcommand{\rb}[1]{\left( #1 \right)}

\newcommand{\ew}[1]{\langle #1 \rangle}
\newcommand{\beq}{\begin{eqnarray}}
\newcommand{\eeq}{\end{eqnarray}}

\newcommand{\op}[2]{| #1 \rangle \langle #2 |}

\newcommand{\eq}[1]{Eq.~(\ref{#1})}
\newcommand{\fig}[1]{Fig.~\ref{#1}}
\newcommand{\secref}[1]{Sec.~\ref{#1}}

\newcommand{\trace}[1]{\mathrm{Tr}\left\{#1\right\}}

\newcommand{\etal}{{\em et al.}\xspace}

\newcommand{\citer}[1]{{Ref.~\cite{#1}}}

\begin{document}
\title{Maximum violations of the quantum-witness equality}
\author{Greg Schild and Clive Emary}
\affiliation{
  Department of Physics and Mathematics,
  University of Hull,
  Kingston-upon-Hull,
  HU6 7RX,
  United Kingdom
}
\date{\today}
\begin{abstract}  
  We consider the quantum-witness test of macroscopic realism and derive an upper bound for possible violations of this equality due to quantum mechanics.  The bound depends only on the number of possible outcomes for the blind measurement at the heart of the witness protocol.  Mirroring recent results for the related Leggett-Garg inequality, we show that quantum mechanics can saturate the algebraic bound for large systems. We also establish a connection between the quantum witness and the trace distance between density matrices, and discuss how the quantum witness can be used to obtain a bound on the Hilbert-space dimension of the system under study.
\end{abstract}
\maketitle

\section{Introduction \label{SEC:intro}}

Macroscopic realism, as codified by Leggett and Garg \cite{Leggett1985}, posits that: 1) a system with two or more macroscopically-distinct states open to it will, at every time, actually be in one or another of those states; and 2) it is possible to ascertain which of these states the system is in without disturbing it.  
Leggett and Garg derived a class of inequalities to test whether these assumptions hold true for a given system \cite{Leggett1985}.  Needless to say, neither assumption holds under quantum mechanics, and violations of these inequalities are taken as evidence for quantum-coherent behaviour. These inequalities have been tested for a wide range of quantum-mechanical systems (see \citer{Emary2014} for a review; more recently \cite{Robens2015,White2015}), and violations of the Leggett-Garg inequalities (LGIs) have been observed in line with quantum-mechanical predictions.

The subject of the current paper is a related test of macroscopic realism, the {\em quantum witness} of \citer{Li2012} (described as {\em no-signalling-in-time} in \citer{Kofler2013}). Consider two observables, $A$ and $B$, measured at times $t=0$ and $t>0$ respectively.  Let $\left\{a_i\right\};~i=1,\ldots,M$ be the outcomes of measurement $A$ with probabilities $P(a_i)$, and let $b$ be a particular outcome of measurement $B$.  Based on the joint measurement of these two observables we can construct the probability of obtaining result $b$ in the later measurement as 
\beq
  P'(b) = \sum_{i=1}^{M} P(b\,|a_i)P(a_i)
  \label{EQ:Kol}
  ,
\eeq
with $P(b\,|a_i)$ the conditional probability of finding outcome $b$ at time $t$ given result $a_i$ at time $t=0$.  The notation $P'$ here is to remind us that this probability is determined in a presence of the measurement of $A$.  Since the results of this measurement are discarded,
we describe this $A$ measurement as a {\em blind measurement}.
We can of course determine the probability of outcome $b$ without this prior measurement and we denote this $P(b)$.  The quantum witness is then defined as the difference 
\beq
  W
  \equiv
  \left|
     P(b) - P'(b)
  \right|
  \label{EQ:Wdef}
  .
\eeq
Under the tenets of macroscopic realism, the presence of the blind measurement, if carried out non-invasively, should not affect the subsequent evolution of the system. The two probabilities should thus be the same and we obtain the quantum-witness equality:
\beq
  W = 0 
  \label{EQ:W0}
  .
\eeq
This is equivalent to the statement that the Chapman-Kolmogorov equation \cite{vanKampenbook} applies to the probabilities associated with (non-invasive) measurement outcomes under macroscopic realism.

The quantum witness tests the same underlying assumptions as the LGIs and can similarly be violated by quantum mechanics. 
The key advantage of \eq{EQ:W0} over the LGIs is its simplicity.  The quantum witness was experimentally tested, and violated, in \citer{Robens2015}.  A test of macroscopic coherence in molecular interferometry using the quantum witness has also been proposed \cite{Emary2014b}.

The questions we pose here are: What are the maximum violations of \eq{EQ:W0} under quantum mechanics, and how do they depend on the system size and the type of measurements we make?

The corresponding questions for the LGIs were investigated in \citer{Budroni2014}. The simplest three-term LGIs (see \citer{Emary2014}) have an upper bound of 1 under macroscopic realism.  From the earliest work \cite{Leggett1985} it was known that the maximum value for a two-level quantum system is $\frac{3}{2}$.  This result was later extended to hold for systems of arbitrary system size but with bipartite measurement projectors \cite{Budroni2013}. 
In \citer{Budroni2014}, however, it was found that this bound can be exceeded, even up to the algebraic bound of three, if a more general measurement scheme is adopted.  In an $N$-level system, it was shown that intermediate measurements described by $2 < M \le N$ projectors could produce greater violations of the LGIs than is the case for $M=2$. The greatest violations were found in the ``von Neumann'' measurement limit when $M=N$.

As far as we are aware, little is known about the maximum quantum values of \eq{EQ:Wdef}.  
The central result of the current work is that, in general, we find the maximum violation of the QW equality is given exactly by
\beq
  W^\text{max} = 1-\frac{1}{M}
  \label{EQ:Wmaxintro}
  .
\eeq
This bound depends only on $M$, the number of possible outcomes of the blind measurement, and arises from the maximisation of the entropy of the post-measurement state.
In the limit of large $M$ (which necessitates large system size since $M\le N$), the witness can reach its algebraic maximum of 1. \eq{EQ:Wmaxintro} is an exact result for arbitrary system size $N$ and outcome number $M$. This simple result contrasts strongly with corresponding bounds derived in \citer{Budroni2014} for the LGIs. These results were known only numerically, and then only for small systems, $N \le 9$, and for a particular model in the $N\to \infty$ limit.

In deriving \eq{EQ:Wmaxintro}, we shall show how the quantum witness is related to the trace distance \cite{Nielsen2000}, a measure of distance between two density matrices, recently studied itself as a measure of ``quantumness'' \cite{Shao2015,Miranowicz2015}. Equation~(\ref{EQ:Wmaxintro}) also opens up the possibility of using the quantum witness as a dimension witness \cite{Brunner2008,Brunner2013} for quantum-mechanical systems.

The paper is structured as follows.
In \secref{SEC:QW}, we write down the quantum-mechanical expression for the witness. We then derive \eq{EQ:Wmaxintro} by considering the unitary evolution of pure states in \secref{SEC:max}.  In \secref{SEC:trace}, we connect the quantum witness to the trace distance, and prove that the pure-state bound from the preceding section is indeed an upper bound for all states and evolutions. \secref{sec:dim} discusses the use of the quantum witness as a dimension witness.  In \secref{SEC:systems}, we consider some simple examples and demonstrate how maximum violation may be achieved in practice. We then finish with some discussions.

\section{Quantum witness \label{SEC:QW}}

We begin by writing a quantum-mechanical expression for the witness $W$. We consider an $N$-dimensional quantum system with density matrix $\rho$ at $t=0$.  The evolution of this state from time $t=0$ to time $t=t$ we describe with the completely-positive trace-preserving map $\Phi$.  Let $\Pi^b$ be the projector corresponding to the measurement outcome $b$, not necessarily one-dimensional.  We shall only consider projective measurements in this work. 
For the directly-measured probability, we have then 
\beq
  P(b) = \trace{\Pi^b \Phi\left[\rho\right]}
  .
\eeq
The blind measurement we specify in terms of the projectors $\left\{ \Pi_i^a \right\}$ with $i = 1,\ldots,M$.  These projectors form a complete set, $\sum_{i=1}^M \Pi_i^a = \mathbbm{1}$, but are not necessarily one-dimensional, $2 \le M \le N$.  After the blind measurement, the system is left in the state
\beq
  \sigma = \sum_{i=1}^M \Pi_i^a \rho \Pi_i^a
  .
\eeq
The required blind-measurement probability is then
\beq
  P'(b) &=& 
  \trace{\Pi^b \Phi\left[\sigma\right]}
  .
\eeq
Quantum-mechanically, therefore, the witness reads
\beq
  W =  
  \left|
    \frac{}{}
    \trace{
      \Pi^b 
      \rb{
        \Phi\left[\rho\right]-\Phi\left[\sigma\right]
      }
    }
  \right|
  \label{EQ:Wdiff1}
  .
\eeq

\section{Pure-state violations \label{SEC:max}}

We begin by considering the case of pure states and unitary evolution.  The maximum value of the witness is simple to find under these conditions and, as we show in the next section, the maximum value holds also for the full problem.

Let us assume that $\rho$ is the pure state $\rho=\op{\phi}{\phi}$ and that $\Pi^b$ is the one-dimensional projector $\Pi^b=\op{\psi^b}{\psi^b}$.  We assume that $\Phi$ describes a unitary evolution that can be absorbed into the definition of $\Pi^b$.  In this case
\beq
  W =  
  \left|
    \left|\ew{\psi^b|\phi}\right|^2
    -
    \sum_{i=1}^M 
    \left|\ew{\psi^b|\Pi_i^a|\phi}\right|^2
  \right|
  .
\eeq
If we choose initial and final measurement states to be identical, $\ket{\psi^b}=\ket{\phi}$, we have
\beq
  W =  
  1
  -
  \sum_{i=1}^M  \left[\ew{\phi|\Pi_i^a|\phi}\right]^2
  =
  1-\sum_{i=1}^M [P(a_i)]^2
  \label{EQ:WasPb}
  .
\eeq
Under the constraint, $\sum_{i=1}^M P(a_i)=1$, this is maximised when all the probabilities are equal: $P(a_i)=\frac{1}{M}$. We thus obtain the bound of \eq{EQ:Wmaxintro}.

For a qubit with $M=N=2$, we obtain $W^\text{qubit}_\mathrm{max} = \frac{1}{2}$.   This is the value reported in \citer{Kofler2013} for a Mach-Zehnder interferometer.  Also consistent with this maximum is the value of $W\approx 0.45$ reported for a two-level system in \citer{Li2012}.
When the measurement A consists of just two projectors, $M=2$, the quantum bound for the witness is exactly the same as for the qubit, irrespective of the system size.  
This result is consistent with \citer{Emary2014b} in which a maximum value of $W\approx 0.48$ was reported for an extended system with a dichotomic intermediate measurement
\footnote{ 
  The witness was defined slightly differently in \citer{Emary2014b} such that the maximum value reported here is half that given in the paper. 
}.
With $M=N$ we obtain
\beq 
  W^\text{vN}_\mathrm{max} = 1 - \frac{1}{N}
  \label{EQ:WmaxvN}
  ,
\eeq
which is the maximum violation possible for a given system size $N$.  In analogy with the corresponding result for the LGI \cite{Budroni2014}, we shall call this the von Neumann bound.  It results from a complete collapse of the wave function under measurement A.

\subsection{Entropy \label{SEC:entropy}}

We can gain insight into this result by using the orthogonality of the projectors to rewrite \eq{EQ:WasPb} as
\beq
   W 
  =  
  1
  -
  \trace{
   \sigma^2
  }
  =S_\mathrm{L}(\sigma)
  \label{EQ:SL}
  ,
\eeq
which is the linear entropy of the post-measurement state $\sigma$.   Under the conditions considered in this section then, the maximum value of $W$ arises when the entropy of the post-measurement state is at a maximum. This occurs when $\sigma$ is the completely mixed state $\sigma = N^{-1} \mathbbm{1}$.

As an aside, we note that this form suggests the definition on an {\em entropic quantum witness}, given as the difference in the entropy between the results of the $B$ measurement with and without the blind measurement:
\beq
  W_\text{ent} 
  \equiv
  \left|
    H'(B) - H(B)
  \right|
  .
\eeq
Here $H(B)$ is the Shannon entropy
\beq
  H(B) 
  =
  -
  \sum_i P(b_i) \log P(b_i) 
  ,
\eeq
with $b_i$ the various results of the $B$-measurement, and $P(b_i)$ their probabilities.  $H'(B)$ is defined in the same way but with $P'(b_i)$ replacing $P(b_i)$. 
An analogous generalisation of the Leggett-Garg inequalities in terms of entropies was discussed in Refs.~\cite{Morikoshi2006,Devi2013}.

\section{Connexion to the trace distance \label{SEC:trace}}

The trace distance between two density matrices, $\rho$ and $\sigma$, is defined as \cite{Nielsen2000}
\beq
  D\rb{
    \rho,\sigma
  }
  \equiv \frac{1}{2}
  \trace{
  \left|
    \rho-\sigma
  \right|
  }
  \label{EQ:TDdefn}
  ,
\eeq
where $|A| = \sqrt{A^\dag A}$ is the positive square-root of $A^\dag A$.
This definition is equivalent to the maximisation
\beq
  D\rb{\rho,\sigma} = \underset{\Pi}{\text{max}} \,\trace{\Pi \rb{\rho-\sigma}}
  ,
\eeq
with respect to all projectors, $\Pi$.  Comparing this with \eq{EQ:Wdiff1}, we find that the maximisation of the quantum witness over all possible $B$-measurements gives precisely the trace distance of the two time-$t$ density matrices:
\beq
  \underset{\Pi^b}{\text{max}} 
  \,
  W
    \!\rb{\Pi^b, \Phi[\rho],\Phi[\sigma]
  }
  =
  D\rb{
    \Phi[\rho],\Phi[\sigma]
  }
  .
\eeq
Here we have explicitly written out the relevant arguments of the witness.
This immediately implies that the witness is bounded from above by the trace distance of the two density matrices at time $t$:
\beq
   W \le 
   D\rb{
    \Phi[\rho],\Phi[\sigma]
  }
  \label{EQ:WDfirst}
  .
\eeq
The trace distance therefore provides a convenient way to calculate an upper bound for a specific pair of density matrices. This we can do by finding the eigenvalues $\Lambda_i$ of the Hermitian matrix $\Phi[\rho]-\Phi[\sigma]$ and calculating $D=\sum_i |\Lambda_i|$.

Furthermore, from \eq{EQ:WDfirst} follow some interesting properties.  Firstly, since it is contractive, the trace distance never increases under the evolution described by $\Phi$.  Thus we have
\beq
   W \le 
   D\rb{
    \Phi[\rho],\Phi[\sigma]
  }\le 
   D\rb{
    \rho,\sigma
  }
  ,
\eeq
and the witness is bounded by the trace distance of the density matrices at the time of the $A$ measurement, $t=0$.

We can then use further properties of the trace distance to show that $W_\text{max}$ of the previous section is the maximum violation with a fixed $M$ for all initial states (including mixed states) and all measurement choices.  Let us begin by writing the initial density matrix as a pure state decomposition: $\rho = \sum_n p_n \op{\phi_n}{\phi_n}$.  The measured density matrix is correspondingly $\sigma = \sum p_n \sigma_n$ with $\sigma_n = \sum_i \Pi_i^a \op{\phi_n}{\phi_n} \Pi_i^a $.   We then use the joint convexivity of the trace distance \cite{Nielsen2000} to write 
\beq 
  D\rb{
    \sum_n p_n \op{\phi_n}{\phi_n},\sum_n p_n \sigma_n
  }
  \le 
  \sum_n p_n
  D\rb{
   \ket{\phi_n}, \sigma_n
  }
  \nonumber
  .
\eeq
In a given decomposition, one particular pure state, $\ket{\phi_m}$ say, will give the maximum value of $D\rb{\ket{\phi_m}, \sigma_m }$. Setting $p_n = \delta_{nm}$ therefore maximises the trace distance for this decomposition.  Maximising over all possible decompositions, we obtain
\beq
  W \le 
  \underset{\ket{\phi}}{\text{max}} \,
  D\!\rb{
   \ket{\phi}, \sum_i \Pi_i^a \op{\phi}{\phi} \Pi_i^a
  }
  .
\eeq
An explicit form for the bound results from the double optimisation
\beq
 W^\text{max} 
 &=&  
 \underset{\Pi^b, \ket{\phi}}{\text{max}} \,
 \trace{ \Pi^b
   \rb{
     \op{\phi}{\phi}- \sum_{i=1}^M \Pi_i^a \op{\phi}{\phi} \Pi_i^a
   }
 }
 ,
 \nonumber\\
 &=&
 \underset{\Pi^b, \ket{\phi}}{\text{max}} \,
 \sum_{ii'=1}^M
 \bra{\phi}
   \Pi_i^a\Pi^b\Pi_{i'}^a
 \ket{\phi}
 \rb{
   1 - \delta_{ii'}
 }
 .
\eeq
The state $\ket{\phi}$ can be found by maximising this with respect to the coefficients $\bra{\phi} \Pi_i^a\Pi^b\Pi_i^a \ket{\phi}$, subject to the constraint 
$
  \sum_{ii'} \bra{\phi} \Pi_i^a\Pi^b\Pi_{i'}^a \ket{\phi}
  =
  \bra{\phi}\Pi^b\ket{\phi} 
$.  At maximum, this gives all contributions to the sum equal, and we obtain
\beq
  W^\text{max} =  
  \underset{\Pi^b}{\text{max}} \,\bra{\phi}\Pi^b\ket{\phi}\rb{1-\frac{1}{M}}
  .
\eeq
Finally, this is optimised by choosing $\Pi^b$ to include the state $\ket{\phi}$. The simplest projector that achieves this is, of course, the one-dimensional projector  $\Pi^b=\op{\phi}{\phi}$.  We then recover \eq{EQ:Wmaxintro}.  This shows that the conditions studied in \secref{SEC:max} are indeed those leading to the most general maximisation of $W$.

\section{Dimension witness \label{sec:dim}}

Since the number of projectors must always be less then or equal to the dimension of the Hilbert space, $ M \le N$, we have
$
  W \le 1 -\frac{1}{M} \le 1-\frac{1}{N} 
$.
Therefore, a measurement of the quantum witness allows us to conclude that the dimension of the underlying Hilbert space satisifies
\beq
  N \ge \frac{1}{1-W}
  \label{EQ:dim}
  ,
\eeq
and the quantum witness discussed here functions as a {\em dimension witness} \cite{Brunner2008} for quantum-mechanical systems (for classical systems, \eq{EQ:dim} gives $N \ge 1$, which is trivial).

In \citer{Brunner2013}, Brunner \etal defined a dimension witness that compares measurements on a number of different preparations of the quantum system under investigation.  Formally, this quantity shares some degree of similarity with the quantum witness studied here, and the upper bound for it was found, as here, by using a connexion with the trace distance. 
For Brunner's witness to give useful information about the dimension of the system's Hilbert space, the number of preparations considered must exceed the Hilbert-space dimension.  
In comparison, the quantum witness compares just two density matrices which, in Brunner's construction, would yield no information about the dimension. 
The reason that the QW is capable of acting as a dimension witness is that the two density matrices in question are not arbitrary, but rather precisely related to one another through the act of the witness' blind measurement.  
Thus, in only comparing two, albeit related, density matrices the QW appears as a rather simple and efficient dimension witness.

\section{Example systems \label{SEC:systems}}

In this section we consider some simple examples of how this upper bound may (and may not) be achieved in practice. As we are interested in maximum violations, we consider pure-state evolution and the von Neumann limit $M=N$.

\subsection{Precessing spin\label{SEC:spin}}

Consider a spin of length $j=N/2$ in a static magnetic field with Hamiltonian
\beq
  H = \Omega \rb{ J_x\cos \theta + J_z \sin \theta}
  .
\eeq
Here, $\Omega$ sets the energy scale ($\hbar=1$) and $\theta$ is an angle that describes the orientation of the field.  The corresponding unitary time-evolution operator is $U(t) = e^{-i H t}$. 
We assume that measurements are restricted to the $J_z$ basis with projectors $\Pi_m = \op{m}{m}$, where $\ket{m}$ is shorthand for the state $\ket{j;m}$; $m=-j,\ldots,+j$.  We assume that the final measurement projector is $\Pi^b =  \op{-j}{-j}$.  The system is prepared in state $\ket{-j}$ at a time $t=-\tau$, such that the state just prior to the blind measurement is $\rho = U(\tau)\op{-j}{-j} U^\dag(\tau)$.  The witness for von Neumann measurement then reads 
\beq
  W 
  &=& 
  \Bigg |
    |\ew{-j|U(2\tau)|-j}|^2
  \nonumber\\
  &&
  ~~~~~~
    -
    \sum_{m=-j}^{+j}
    |\ew{
    -j|
    U(\tau)
    \Pi_m
    U(\tau)
    |-j
    }|^2
  \Bigg |
  .
\eeq

The $j=\frac{1}{2}$ case (qubit) is straightforward.  We set $\theta =0$ in the Hamiltonian to obtain
\beq
  W^\text{qubit} = \textstyle{\frac{1}{2}} \sin^2( \Omega t)
  .
\eeq
This has a maximum value of $\textstyle{\frac{1}{2}}$, as given by \eq{EQ:WmaxvN}, when $2 \Omega t$ is an integer multiple of $\pi$.
In obtaining this maximum, it is the first term in the witness that is exactly zero.  This illustrates that the maximum can also be found with orthogonal $\ket{\psi}$ and $\ket{\phi}$.

\begin{figure}[t]
  \begin{center}
    \includegraphics[width=\columnwidth,clip=true]{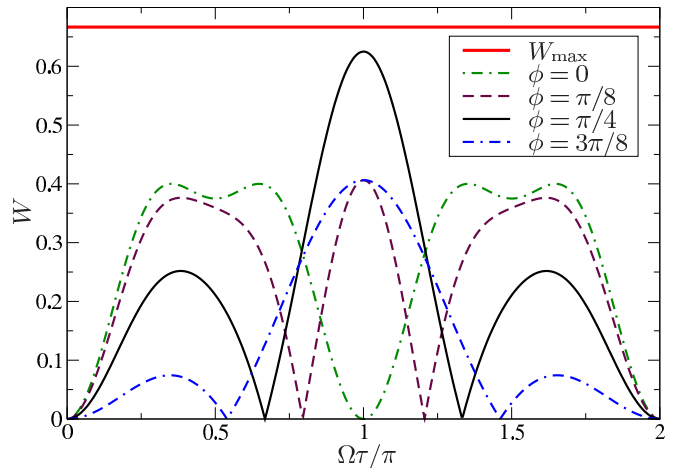}   
  \end{center}
  \caption{
   The value of quantum witness $W$ for a precessing spin of length $j=1$ as a function of time $\tau$. Results for several values of the angle
   $\theta$ are shown.  The overall maximum value is obtained by setting $\theta = \pi/4$ and $\Omega \tau = \pi$.  This gives $W = 5/8$, which is greater than the qubit value of one-half, but less than the maximum value $W_\text{max} = 2/3$ predicted by \eq{EQ:WmaxvN} for $N=M=3$ (indicated by the horizontal line).
    \label{FIG:QW3LS}
  }
\end{figure}

The results for $j=1$ are a bit more involved and these are shown in \fig{FIG:QW3LS}.
The value of the witness depends on the direction of the magnetic field.  The maximum violation occurs for $\theta = \pi/4$ and at a time $\Omega t =\pi$, where it takes a value of  $W=5/8$. This is, however, less than the value of $W=2/3$ predicted by \eq{EQ:WmaxvN}.  Neither adding a $J_y$ component to the Hamiltonian, nor changing the relative duration of the two time evolutions serves to increase this value. Thus, subject these dynamics and measurements, the value $W=5/8$ is the best that can be achieved for the precessing spin.

\subsection{Controlled evolution\label{SEC:cdyn}}

\begin{figure}[t]
  \begin{center}
    \includegraphics[width=\columnwidth,clip=true]{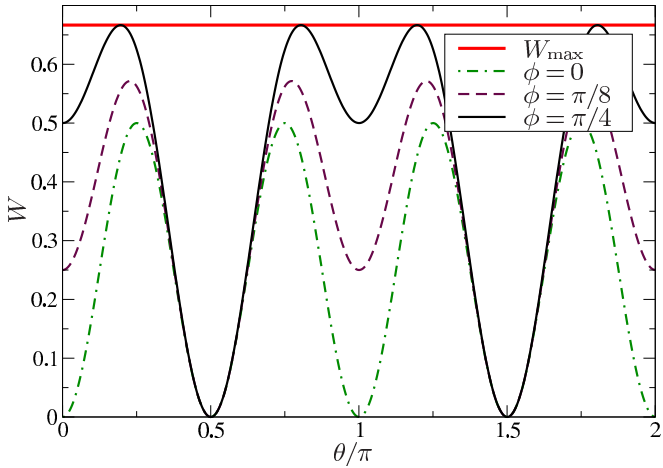}   
  \end{center}
  \caption{
   Value of the quantum witness for a $j=1$ system  with controlled evolution described by the two angles $\theta$ and 
   $\phi$.  The maximum value found here is $W = 2/3$, which saturates the upper bound of \eq{EQ:WmaxvN}.
    \label{FIG:QW3LSU}
  }
\end{figure}

The above failure to saturate the bound can be seen as a result of the  restriction of the time-evolution operator $U$ to SU(2) rotations.  Let us therefore consider a second scenario in which the quantum dynamics can be chosen freely.  We consider two different evolutions: one from $t=-\tau$ to $t=0$, and then one from $t=0$ to $t=\tau$, and design the second to be the inverse of the first.  This is reminiscent of a Ramsey experiment (see Refs.~\cite{Asadian2014,Robens2015} for connections with tests of macrorealism) with two $\pi/2$ pulses returning the system to the initial state. The variable time delay of the Ramsey experiment, however, is replaced here by either the  presence or absence of the $A$ measurement.  With this choice the first term in the witness is equal to unity, and we have
\beq
  W = 
  \left|
    1
    -
    \sum_{m=-j}^{+j}
    |\ew{
    -j|
    U^\dag
    \Pi_m
    U
    |-j
    }|^2
  \right|
  \label{EQ:WRamsey}
  .
\eeq
Considering the $j=1$ spin again, let us choose the unitary evolution operator to be
\beq
  U = 
  \rb{
  \begin{array}{ccc}
    \cos \theta & 0 & \sin \theta \\
    \sin \theta \sin \phi & \cos\phi & -\cos \theta \sin \phi\\
    -\sin \theta \cos \phi& \sin \phi & \cos \theta\cos \phi
  \end{array}
  }
  .
\eeq
This gives the witness as
\beq
  W = |
    1 - \textstyle{\frac{1}{4}} \rb{3 + \cos 4 \phi} \cos^4 \theta - \sin^4\theta
  |
  ,
\eeq
a result plotted in \fig{FIG:QW3LSU}. The maximum value of this expression occurs at e.g. $\theta = \arccos \sqrt{2/3}$ and $\phi = 3\pi/4$, where is takes the value of $W=2/3$.  This saturates the bound of \eq{EQ:WmaxvN} for $N=M=3$.

\subsection{Algebraic limit \label{SEC:alg}}

\eq{EQ:WmaxvN} makes it clear that in the $(N,M)\to \infty$ limit, we should be able to achieve the algebraic bound for the quantum witness.  We now give a concrete example of how this might be achieved.

Consider a single bosonic mode with Fock states $\ket{n}$ and let the evolution operator from time $t=-\tau$ to $t=0$ be the displacement operator $\mathcal{D}(\alpha)$  with displacement parameter $\alpha$, and that from $t=0$ to $t=t$ be its inverse $\mathcal{D}(-\alpha)$.
With initial state and final projector equal to the vacuum state $\op{0}{0}$, the important matrix elements of the displacement operator are \cite{Cahill1969}
\beq
  \ew{n|
    \mathcal{D}(\alpha)
  |0}
  =
  \frac{1}{\sqrt{n!}} \alpha^n e^{- |\alpha|^2/2}
  .
\eeq
From an expression analogous to \eq{EQ:WRamsey}, we obtain a value for the witness
\beq
  W = 1 - e^{-2|\alpha|^2} I_0(2 |\alpha|^2)
  ,
\eeq
with $I_0(z)$ a modified Bessel function of the first kind. For large $z$, this has the asymptotic form $I_0(z) \sim e^z/\sqrt{2 \pi z}$, such that, for large displacements we obtain
\beq
  W \sim 1 - \frac{1}{2 \sqrt{\pi} |\alpha|}
  .
\eeq
Clearly this gives the algebraic bound $W = 1$ in the limit $|\alpha| \to \infty$.  This may at first seem surprising as $|\alpha|\to \infty$ is usually thought of as the classical limit for coherent states, and yet the witness reaches in ``most quantum'' value in this limit. This is resolved by noting that the measurement here is of an extreme quantum nature, and resolves the coherent states down to  the level of the individual Fock states.

As discussed in \citer{Budroni2014}, the evolution of a bosonic mode under displacements can be used to approximate to the dynamics of the spin model of section \secref{SEC:spin} in the large-spin limit and with $\theta=0$.  This approximation identifies $ \alpha = \sqrt{j/2} \,\Omega t$. The algebraic bound is therefore attainable in the large-spin limit of the precessing spin.  The difference between the witness and its algebraic limit is seen to scale as $j^{-1/2}$. This is the same behaviour as was observed for the LGI applied to this model.

\section{Discussion\label{SEC:disc}}

Our main result is the explicit bound for the quantum witness, \eq{EQ:Wmaxintro}, which depends only on the number of different outcomes for the blind measurement.   Since this number is naturally limited by the number of quantum-mechanical levels possessed by the system, our result shows that, the larger the system, the greater the possible violation of the witness.
Fundamentally it is the greater information gained about the system by the measurement at $t=0$,  and the corresponding greater ``collapse'' of the wave function, that is responsible for these increased violations.  This is reflected in \eq{EQ:SL}, which equates the value of the witness under ideal conditions to the linear entropy of the system after the blind measurement.

Qualitatively these results mirror those known for the LGI.  The advance of the current study is that, since the quantum witness is a significantly simpler object than the Leggett-Garg correlators, the bounds here have been derived analytically for arbitrary $N$ and $M$. 

We have also determined the relationship between the quantum witness and the trace distance. This we used here in establishing the upper bound for $W$.  This connection should also be useful in calculating the maximum violation possible for a given quantum evolution, as well as  determining the measurements required to obtain this value.  
This result also ties the experimentally-testable quantum witness to more formally-defined measures of ``quantumness'' \cite{Girolami2014,Miranowicz2015}.  Indeed,  the trace norm itself has recently been considered as a measure of quantumness \cite{Shao2015,Miranowicz2015}, were the density matrix under scrutiny is compared with a particular set of classical states.   The quantum witness studied here can thus be seen as an indicator of this class in which the comparator state is the very specific state that results from the witness' blind measurement. 
Finally, we note that this comparison of a state and its measured counterpart enables the quantum witness to act as a particularly-simple dimension witness.

\begin{acknowledgments}
  We are grateful to Neill Lambert for helpful discussions.
\end{acknowledgments}


\end{document}